\documentclass[aps,prl,amsmath,amssymb,twocolumn,floatfix,superscriptaddress,nofootinbib]{revtex4}
\pdfoutput=1
\usepackage{graphicx}
\usepackage{amsmath, amsthm, amssymb}
\usepackage{slashed}
\usepackage{url}
\usepackage[pdftex]{hyperref}
\usepackage{color}
\usepackage{bm}
\usepackage[utf8]{inputenc} 
\usepackage[normalem]{ulem}

\newcommand{\squark}{\tilde{q}}
\newcommand{\bino}{\tilde{B}}
\newcommand{\bivo}{bi$\nu$o~}

\begin{document}
\vspace*{-1.7cm}

\begin{flushright}{\small FERMILAB-PUB-16-225-T}\end{flushright}

\title{Neutrino masses from a pseudo-Dirac Bino} 
\author{ Pilar Coloma}\email{pcoloma@fnal.gov}
\author{Seyda Ipek}\email{sipek@fnal.gov}
\affiliation{Fermi National Laboratory, Batavia, IL 60510, USA}

\date{\today}
\begin{abstract}
We show that, in $U(1)_R$-symmetric supersymmetric models, the bino and its Dirac partner (the singlino) can play the role of right-handed neutrinos and generate the neutrino masses and mixing, without the need for traditional bilinear or trilinear R-parity violating operators. 
The two particles form a pseudo-Dirac pair, the `\bivo\!\!'. An Inverse Seesaw texture is generated for the neutrino-\bivo sector, and the lightest neutrino is predicted to be massless. Unlike in most models with heavy right-handed neutrinos, the \bivo can be sizably produced at the LHC through its interactions with colored particles, while respecting low energy constraints from neutrinoless double-beta decay and charged lepton flavor violation.
\end{abstract}
\maketitle

Neutrino oscillations have revealed that at least two of the neutrinos in the Standard Model (SM) are massive. While their absolute mass scale is not yet known, neutrino squared-mass differences ($\Delta m^2_{ij} \equiv m_i^2 - m_j^2$) and mixing angles are measured to be~\cite{Gonzalez-Garcia:2014bfa}
\begin{align}
&\Delta m_{21}^2\simeq 7.6\times 10^{-5}~{\rm eV}^2,\quad |\Delta m_{31}^2|\simeq 2.4\times 10^{-3}~{\rm eV}^2,\notag \\
&\sin^2\theta_{12}\simeq 0.3,\quad \sin^2\theta_{23}\simeq 0.47,\quad \sin^2\theta_{13}\simeq 0.024 . \label{eq:nuparam}
\end{align}

It is not possible to generate the neutrino masses within the SM particle content with renormalizable operators and, thus, new physics is needed. The most straightforward avenue is through the addition of right-handed (RH) neutrinos and a Yukawa coupling $Y_\nu$ to the Higgs and the left-handed neutrinos. In Seesaw Type-I models~\cite{Minkowski:1977sc,GellMann:1980vs}, the smallness of the light neutrino masses is explained via a suppression by the RH neutrino Majorana mass $M_R$, as $m_\nu \sim Y_\nu^{\rm T} M_R^{-1} Y_\nu v^2$, where $v$ is the Higgs vacuum expectation value (vev). For instance, for $M_{R}\sim O(10^{14}~{\rm GeV})$, and $Y_\nu \sim \mathcal{O}(1)$, light neutrino masses are at the eV scale. In these models, lepton number is violated in the Lagrangian through the large RH Majorana masses. 

Another way to account for the smallness of neutrino masses is the Inverse Seesaw (ISS) mechanism~\cite{Mohapatra:1986aw,Mohapatra:1986bd,Branco:1988ex}. In this case, RH neutrinos form pseudo-Dirac pairs and lepton numbers are assigned in a way that the total lepton number is (approximately) conserved in the Lagrangian. In this framework, all lepton number violating effects (such as light neutrino masses) are naturally suppressed. This allows to lower the RH neutrino mass scale, for example to the TeV scale, without introducing very small Yukawa couplings. The phenomenological consequences in these scenarios are wide, as they can lead to observable rates in lepton flavor violating processes, or at the LHC. The challenge, however, is to explain the texture needed in the neutrino mass matrix, \textit{i.e.}, the pairing of the heavy neutrinos in pseudo-Dirac states and the suppression of lepton number violating effects. 

In this letter, we focus on a supersymmetric extension of the SM, and construct a natural realization of an ISS scenario which can fully account for the observed neutrino masses and mixing at low energies. Most supersymmetric models that address this problem typically generate neutrino masses via R-parity violating (RPV) bilinear terms of the form $\mu_i H_u L_i$ (see, \emph{e.g.},~\cite{Barbier:2004ez,Hirsch:2004he} and references therein). In these models, however, it is usually necessary to include trilinear RPV couplings and/or RH neutrinos to explain the neutrino mixing structure\footnote{This is not always the case, see \emph{e.g.}, Refs.~\cite{Hirsch:2004he, Abada:2006qn,Abada:2006gh}. }. Here we follow a different approach. In $U(1)_R$-symmetric supersymmetric models with Dirac gauginos, the bino and its Dirac partner are gauge singlets of the SM and therefore can play the role of the RH neutrinos. Furthermore, they automatically form a pseudo-Dirac pair and lead to an ISS texture for the neutrino mass matrix. Small gaugino Majorana masses are generated by $U(1)_R$-violating effects, namely the gravitino mass. We show that this framework can explain neutrino masses and mixing without any bilinear or trilinear RPV terms. More importantly, \emph{no new RH neutrinos} are needed. Finally, unlike in traditional ISS scenarios, in this model the RH ``neutrinos'' can be produced in decays of colored particles. Thus, sizable production cross sections can be obtained at the LHC without being in conflict with low-energy observables, such as neutrinoless double-beta decay or $\mu \rightarrow e \gamma$ constraints. 

\emph{The Model.}
In $U(1)_R$-symmetric supersymmetric models~\cite{Hall:1990hq, Kribs:2007ac}, superpartners have $+1$ $R$--charges while the SM fields are not charged under $U(1)_R$. Gaugino Majorana masses are thus forbidden. In order to give Dirac masses to gauginos, adjoints with opposite $R$-charges are introduced. Dirac gauginos alleviate supersymmetric CP and flavor problems and require less fine-tuning for large gluino masses~\cite{Kribs:2007ac, Fox:2002bu}. In these scenarios, Higgsino masses are also forbidden; however they can be generated by extending the Higgs sector. A 125 GeV Higgs mass can also be accommodated, see \emph{e.g.},~\cite{Fok:2012fb,Bertuzzo:2014bwa}. Here we follow a similar approach, but instead of a $U(1)_R$ symmetry we consider a global $U(1)_{R-L}$ symmetry, where $L$ is the lepton number, as discussed in Refs.~\cite{Frugiuele:2011mh,Bertuzzo:2012su}\footnote{It should be stressed that while in Ref.~\cite{Bertuzzo:2012su} neutrino masses are generated through the bilinear and trilinear RPV terms, these will not be considered here. }. For convenience, Table~\ref{table:fields} lists the $U(1)_{R-L}$ charges of the  fields relevant for this study. 
\begin{table}[htb!]
\begin{tabular}{|c|c|c|c|c|c|}
\hline
Superfields	&	$SU_c(3)$	&	$SU_L(2)$	&	$U_Y(1)$	&	$U(1)_R$	&	$U(1)_{R-L}$ \\
\hline\hline
$L_i$	&	1	&	2	&	-1/2	&	1	&	0 \\
$E^c_i$	&	$1$		&	1	&	1	&	1	&	2	\\	
\hline
$H_u$	&	1	&	2	&	1/2	&	0	&	0	\\
\hline
$W_{\tilde{B}}^\alpha$	&	1	&	1	&	0	&	1	&	1	\\
$\Phi_S$	&	1	&	1	&	0	&	0	&		0	\\
\hline
$W'_\alpha$	&	1	&	1	&	0	&	1	&	1	\\
\hline
\end{tabular}
\caption{Superfields relevant for the discussion and their charge assignments. $L_i,E_i^c$ are the lepton superfields and $H_u$ is the up-type Higgs superfield. The subindex $i$ indicates the fermion generation. The fermionic component of $\Phi_S$, $S$, is the Dirac partner of the bino, $\bino$ and is called the singlino. $W'_\alpha$ is a spurion field with a $D$ term. The $R-L$ charge is obtained as the $U(1)_R$-charge minus the lepton number of the field.} \label{table:fields}
\end{table}

We assume that supersymmetry is broken in a hidden sector that communicates with the visible sector at a messenger scale $\Lambda_M$, and supersymmetry breaking is incorporated via the spurion field $W'_\alpha = \theta_\alpha D$, where $D$ is a supersymmetry-breaking order parameter which is $U(1)_{R-L}$-neutral. A Dirac mass for the bino is generated as~\cite{Fox:2002bu}
\begin{equation}
\int d^2\theta\, c\frac{W'_\alpha}{\Lambda_M}W_{\bino}^\alpha\Phi_S
\to \frac{cD}{\Lambda_M}\bino S\equiv M_D\bino S, \label{eq:MD}
\end{equation}
where $c$ is a dimensionless coefficient of $\mathcal{O}(1)$ and $\Phi_S$ is the chiral superfield whose fermionic component is the singlino $S$, \textit{i.e.}, the Dirac partner of the bino. Specifically, $\bino$ and $S$ are the Weyl components of the Dirac field $\psi^{\rm T}=(\bino,~S^\dagger)$. The (pseudo)scalar component of $\Phi_S$ can get a negative contribution to its mass squared from the supersoft term $W'_\alpha W'^\alpha \Phi_S \Phi_S$. These terms however can be forbidden, see \emph{e.g.}, Ref.~\cite{Alves:2015kia}.

$U(1)_{R-L}$ must be broken by supergravity. Then Majorana masses for the bino and the singlino will be generated through anomaly mediation~\cite{Randall:1998uk,Giudice:1998xp}, 
\begin{align}
m_{\tilde{B}} \sim m_{S} \sim \frac{1}{16\pi^2}m_{3/2} \, ,
\end{align}
where $m_{3/2}$ is the gravitino mass. We take $m_{3/2}$ to be small, so that $U(1)_{R-L}$ is approximately conserved. Thus, for $M_D \gg m_{3/2}$ the bino and the singlino acquire small Majorana masses and $\psi^{\rm T}=(\bino,~S^\dagger)$ becomes a \emph{pseudo-Dirac} fermion, which will be referred to as the  `\bivo\!\!' in the rest of this work. 

The phenomenology of $U(1)_R$--symmetric supersymmetric models has been extensively studied in the literature, see, \emph{e.g.}, Refs.~\cite{Kribs:2013oda, Dudas:2013gga, Diessner:2014ksa}. Here we focus on the \bivo interactions that are relevant for generating the light neutrino masses and mixing. First let us consider the $U(1)_{R-L}$--conserving $d=6$ operator
\begin{equation}
\frac{f_i}{\Lambda_M^2}\int d^2\theta\, W'_\alpha W^\alpha_{\bino} H_u L_i \to \frac{f_i' M_D}{\Lambda_M}h_u\bino\ell_i~. \label{eq:Bnu}
\end{equation}
Here, $f_i $ and $f'_i \equiv f_i/c$ are dimensionless coefficients of $\mathcal{O}(1)$, and the index $i$ refers to the lepton family. After electroweak symmetry breaking (EWSB), this term generates a Dirac mass term between the active neutrinos and $\tilde{B}$. In this sense, \textit{the bino acts as a heavy right-handed neutrino in this model.} Note that although this term violates R-parity and lepton number, it conserves $U(1)_{R-L}$.

The interaction above is not enough to explain the neutrino oscillation data, though, since there are not enough degrees of freedom to give different masses to (at least) two of the SM neutrinos. However, the singlino can also contribute to the generation of light neutrino masses through an analogous term to that in Eq.~\ref{eq:Bnu}. The operator $\Phi_S H_u L_i$ is not charged under $U(1)_{R-L}$ and cannot be present in the superpotential. Nevertheless, it can be introduced through a $d=5$ K\"{a}hler potential term using the conformal compensator $\phi=1+\theta^2m_{3/2}$,
\begin{align}
\int d^2\theta d^2\bar{\theta}\,\phi^\dagger\,\frac{d_i \Phi_S H_u L_i}{\Lambda_M}~, \label{eq:conformal}
\end{align}
where $d_i$ are dimensionless coefficients. Eq.~\ref{eq:conformal} leads to the following $U(1)_{R-L}$--breaking contribution to the superpotential:
\begin{align}
\frac{m_{3/2}}{\Lambda_M}d_i\int d^2\theta\, \Phi_S H_u L_i\to \frac{d_i m_{3/2}}{\Lambda_M} h_u S\ell_i ~. 
\label{eq:Snu}
\end{align}
Therefore, \textit{the singlino acts as the second right-handed neutrino.} We emphasize that the coupling in Eq.~\ref{eq:Snu} is highly suppressed with respect to the one in Eq.~\ref{eq:Bnu} as it violates $U(1)_{R-L}$, with $ m_{3/2} \ll M_D$. Next we will show the interactions in Eqs.~\ref{eq:Bnu} and \ref{eq:Snu} alone are able to explain the neutrino masses and mixing.  

\emph{Neutrino Masses.}
For simplicity, we take the lightest neutralino to be a pure \bivo \footnote{A significant mixing with the higgsino and the wino will just modify the neutralino mixing matrix, but will not affect the neutrino sector.}. We also assume that the \bivo is the lightest supersymmetric particle (LSP), besides the gravitino. Using the interactions in Eqs.~\ref{eq:Bnu} and \ref{eq:Snu}, in the basis $(\nu_i,~\tilde{B},~S)$, the neutrino-\bivo mass matrix is \footnote{The up- and down-type Higgs field vevs, $v_u$ and $v_d$ respectively, satisfy $v^2=v_u^2+v_d^2=(246~{\rm GeV})^2$. Using $\tan\beta=v_u/v_d$, we have $v_u^2=\frac{v^2}{1+(\tan\beta)^{-2}}$. For $\tan\beta\gg 1$, we use $v_u\simeq v$. It is straightforward to re-derive the bounds in this paper for different values of $\tan\beta$.}
\begin{align}
\mathbb{M}=\left(\begin{array}{c c c}
			\mathbf{0}_{3\times 3}	&	\mathbf{Y} v	&	\mathbf{G}\, v\\
			\mathbf{Y}^{\rm T} v	&	m_{\bino}	&	M_D\\
			\mathbf{G}^{\rm T} v	&	M_D	&	m_S\\
			  \end{array}\right), \label{eq:Mnu}
\end{align}
where $\mathbf{Y}^{\rm T}=(Y_e,~Y_\mu,~Y_\tau)$ and $\mathbf{G}^{\rm T}=(G_e,~G_\mu,~G_\tau)$ are generated through Eqs.~\ref{eq:Bnu} and~\ref{eq:Snu} after EWSB, as $Y_i \equiv f'_i M_D/\Lambda_M$ and $G_i \equiv d_i m_{3/2}/\Lambda_M $. Therefore, a large hierarchy between $Y$ and $G$ is naturally expected, due to the hierarchy $m_{3/2} \ll M_D$. As already mentioned, the hierarchy $m_{\bino}, m_{S} \ll M_D$ is also expected. Hence, the above mass matrix automatically assumes an ISS texture.

A detailed diagonalization of the neutrino mass matrix for an ISS texture has been performed, \emph{e.g.}, in Ref.~\cite{Gavela:2009cd}. We focus on a normal ordering scenario, \textit{i.e.}, $\Delta m^2_{31}>0$. It is straightforward to generalize the results for the inverted ordering scenario, $\Delta m^2_{31} < 0$. In order to recover the correct mixing structure, $Y_i$ and $G_i$ should have the form~\cite{Gavela:2009cd}
\begin{align}
\begin{array}{ccc}
Y_i &=&\dfrac{M_D}{\sqrt{2}\Lambda_M }\left( \sqrt{1+\rho}\,\,U_{i3}^\ast+\sqrt{1-\rho}\,\,U_{i2}^\ast \right), \\[4mm]
G_i &=&\dfrac{m_{3/2}}{ \sqrt{2}\Lambda_M}\left( \sqrt{1+\rho}\,\,U_{i3}^\ast-\sqrt{1-\rho}\,\,U_{i2}^\ast \right), 
\label{eq:YG}
\end{array}
\end{align}
where $\rho\simeq 0.7$ is determined by the neutrino mass splittings, and $U$ is the light neutrino mixing matrix. We ignore the two possible $CP$-violating phases in this system as they do not affect the discussion significantly. With the mixing parameters given in Eq.~\ref{eq:nuparam} we get
\begin{align}
\mathbf{Y}\simeq \frac{M_D}{\Lambda_M}\left(\begin{array}{c}
		0.35 \\
		0.85\\
		0.39 \end{array}\right),\quad \mathbf{G}\simeq \frac{m_{3/2}}{\Lambda_M}\left(\begin{array}{c}
											-0.06 \\
											0.44\\
											0.89 \end{array}\right).
\end{align}
In finding the above relations we ignore the Majorana masses. Non-zero Majorana masses modify $G\to G+\frac{m_S}{M_D}Y$. Thus they can be ignored as long as $G_i \gg Y_i m_S/M_D$, which is always satisfied in this model since $m_S\simeq m_{3/2}/(16\pi^2)\ll m_{3/2}$.

The mass matrix $\mathbb{M}$ has one zero eigenvalue corresponding to a massless neutrino. The other two light neutrino masses are given by
\begin{align}
m_2=\frac{m_{3/2}\,v^2}{\Lambda_M^2}(1-\rho), \quad
m_3=\frac{m_{3/2}\,v^2}{\Lambda_M^2}(1+\rho).
\label{eq:numass}
\end{align}
We emphasize that the neutrino masses in Eq.~\ref{eq:numass} are independent of the Dirac \bivo mass, unlike in most neutrino mass models. Hence $M_D$ is still a free parameter in the model. However, in the discussion above, neutrino masses are obtained in an effective operator approach, which implicitly assumes that $M_D \gg v$. We will therefore take the \bivo to be at the TeV scale. For a benchmark value of $\Lambda_M=100~$TeV, this requires $\sqrt{D}\simeq10~$TeV, see Eq.~\ref{eq:MD}. Taking $m_{3/2}\sim\mathcal{O}(\rm{keV})$ as our benchmark value\footnote{While the gravitino can decay into neutrinos and photons via the neutrino-\bivo mixing, its lifetime $\Gamma^{-1}(\tilde{G}\to \nu\gamma)\simeq M_{\rm pl}^2/(\theta^2 m_{3/2}^3) \simeq 10^{39}~{\rm s}$ is long enough to be a good dark matter candidate~\cite{Takayama:2000uz}. Here $\theta\sim 10^{-3}$ is the neutrino-\bivo mixing angle, see later text.}, the neutrino mass constraints require a messenger scale $\Lambda_M \sim \mathcal{O}(10-100~{\rm TeV})$, in order to reproduce the mass splittings in Eq.~\ref{eq:nuparam}.  Hence, in this model the (assumed) hierarchy between $\Lambda_M$ and $m_{3/2}$ directly relates to the hierarchy between neutrino masses and the EW scale. 

\emph{Lepton Flavor Violation.} Lepton flavor violating (LFV) observables severely constrain $Y_i$ in Eq.~\ref{eq:YG} (see, e.g., Refs.~\cite{Gavela:2009cd,Antusch:2006vwa,Antusch:2014woa,Fernandez-Martinez:2016lgt}). Current upper bounds on LFV decays of charged leptons are~\cite{TheMEG:2016wtm, Aubert:2009ag, Hayasaka:2007vc}
\begin{align}
{\rm Br}(\mu\to e \gamma)&<4.2\times 10^{-13}~,\notag \\
{\rm Br}(\tau\to e \gamma)&<3.3\times 10^{-8}~,  \\
{\rm Br}(\tau\to \mu \gamma)&<4.4\times 10^{-8}~. \notag
\end{align}

The strongest limit comes from $\mu\to e \gamma$ and implies
\begin{align}
\frac{v^2}{2M_D^2}Y_eY_\mu^\ast<2.4\times 10^{-5}\Longrightarrow\Lambda_M\gtrsim30~{\rm TeV}.
\end{align}
Additional limits can be obtained from $\mu\to e$ conversion in nuclei. Although these are not yet as strong, future experiments like Mu2e~\cite{Bartoszek:2014mya} are expected to further constrain $\Lambda_M$. Following Ref.~\cite{Alonso:2012ji}, we obtain a projected sensitivity for the Mu2e experiment of $\Lambda_M \gtrsim 65~ \rm{TeV}$.

\begin{figure} 
\includegraphics[width=.9\linewidth]{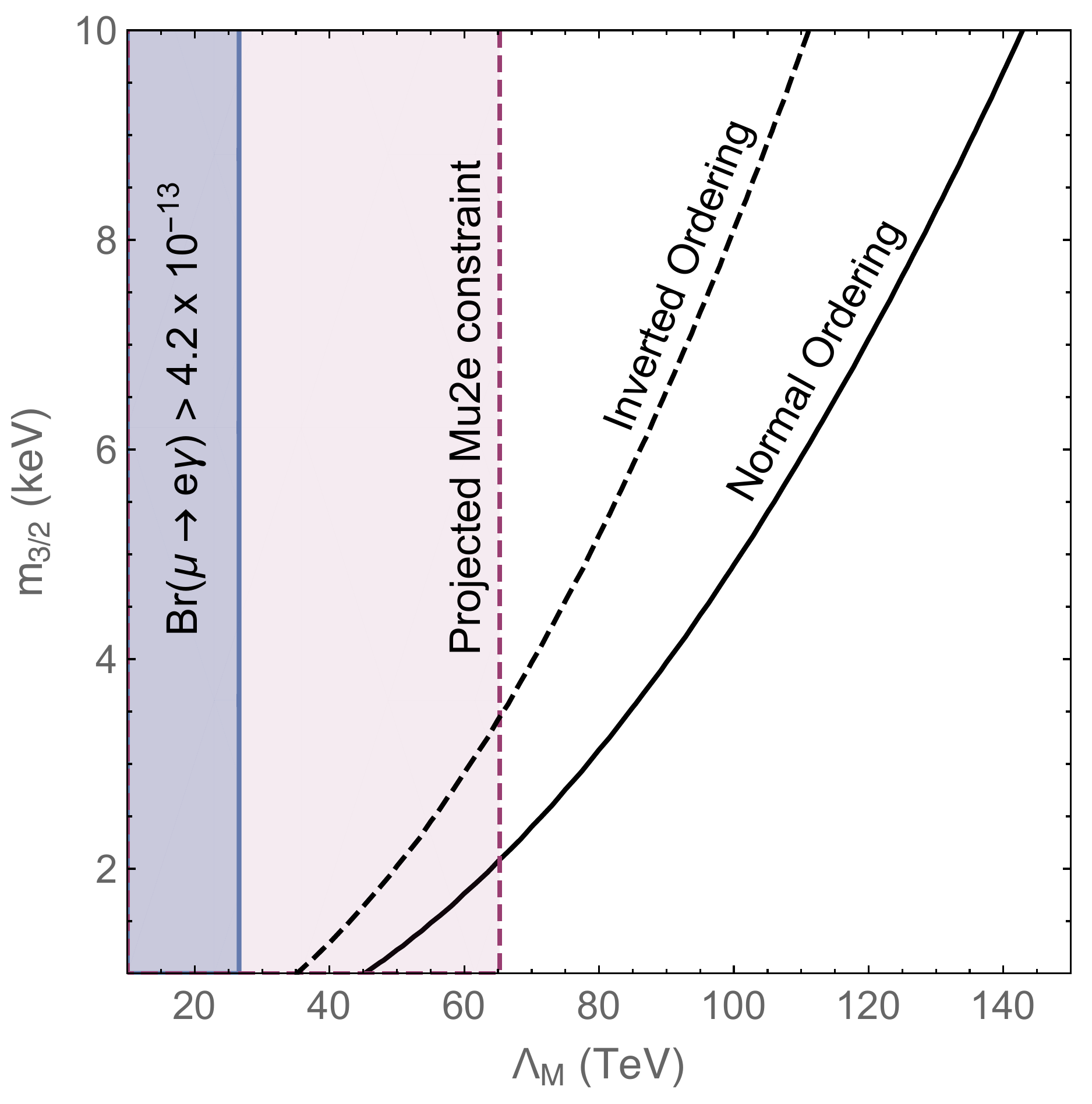}
\caption{Constraints on the messenger scale $\Lambda_M$ and the gravitino mass $m_{3/2}$. The dark shaded region is excluded by $\mu\to e\gamma$ searches~\cite{TheMEG:2016wtm}. The black curves indicate the values needed to reproduce the light neutrino masses for a normal (solid) and inverted (dashed) neutrino mass ordering. In the light shaded region we show the future reach from $\mu\to e$ conversion in nuclei, projected by the Mu2e experiment~\cite{Bartoszek:2014mya}.} \label{fig:Lambda-m32}
\end{figure}

We conclude this section by combining the constraints from LFV observables and neutrino masses in Fig.~\ref{fig:Lambda-m32}. The solid and dashed lines indicate the values of $m_{3/2}$ and $\Lambda_M$ that can reproduce the right neutrino mixing and masses at low energies in this model, for the normal and inverted ordering scenarios. The shaded area shows the region ruled out by the $\mu \to e \gamma$ constraint (the projected reach from $\mu\to e$ conversion in nuclei is also shown). As can be seen from this figure, the combination of the two translates into a lower bound on the gravitino mass, $m_{3/2}\gtrsim {\rm keV}$.

\emph{Neutrinoless double-beta decay.} 
In the ISS limit ($M_D \gg m_B, m_S $ and $ M_D, Yv \gg Gv $), the additional contributions to the neutrinoless double-beta decay rate due to the exchange of heavy neutrinos can be expressed as~\cite{Lopez-Pavon:2015cga}
\begin{eqnarray}
m_{0\nu\beta\beta}^{heavy} &\sim & f(A)\frac{\Lambda_A^2 v^2 }{2M_D^4}\left( (2 m_B + m_S ) Y_{e}^2 - 2 M_D Y_{e} G_{e}\right) \nonumber \, \label{eq:0nubbH},
\end{eqnarray}
where $f(A) \sim\mathcal{O}(0.1)$, and $\Lambda_A \sim 0.9\;\rm{GeV}$ come from an approximation for the nuclear form factor~\cite{Blennow:2010th,Ibarra:2010xw}. This contribution is largely suppressed and the rate is  below the current constraints, $m_{0\nu\beta\beta} < 60$~meV~\cite{KamLAND-Zen:2016pfg}. 

In principle, corrections at 1-loop could induce neutrinoless double-beta decay directly through a non-zero $e-e$ entry in the light neutrino mass matrix~\cite{Pilaftsis:1991ug,Grimus:2002nk,Dev:2012sg}. However, these corrections are also strongly suppressed since they are proportional to~\cite{Lopez-Pavon:2015cga} $ 1/(4\pi)^2\theta^2 m_B \sim \mathcal{O} \left( m_{3/2} v^2/((4\pi)^4 \Lambda_M^2)\right) $, where $\theta \sim Yv/M_D$ stands for the mixing between the light and heavy neutrino states.

\emph{Collider Phenomenology.}  Collider bounds for $U(1)_R$-symmetric models tend to be weaker than for other supersymmetric scenarios. For instance, the limits on squark masses in $U(1)_{R-L}$-symmetric models can be as low as 600~GeV~\cite{Frugiuele:2012kp}. A detailed study of current LHC bounds on our model is left for future work. Here we outline the general signatures expected at the LHC for a particular choice of benchmark parameter values. 

\begin{figure} [th!]
\includegraphics[width=\linewidth]{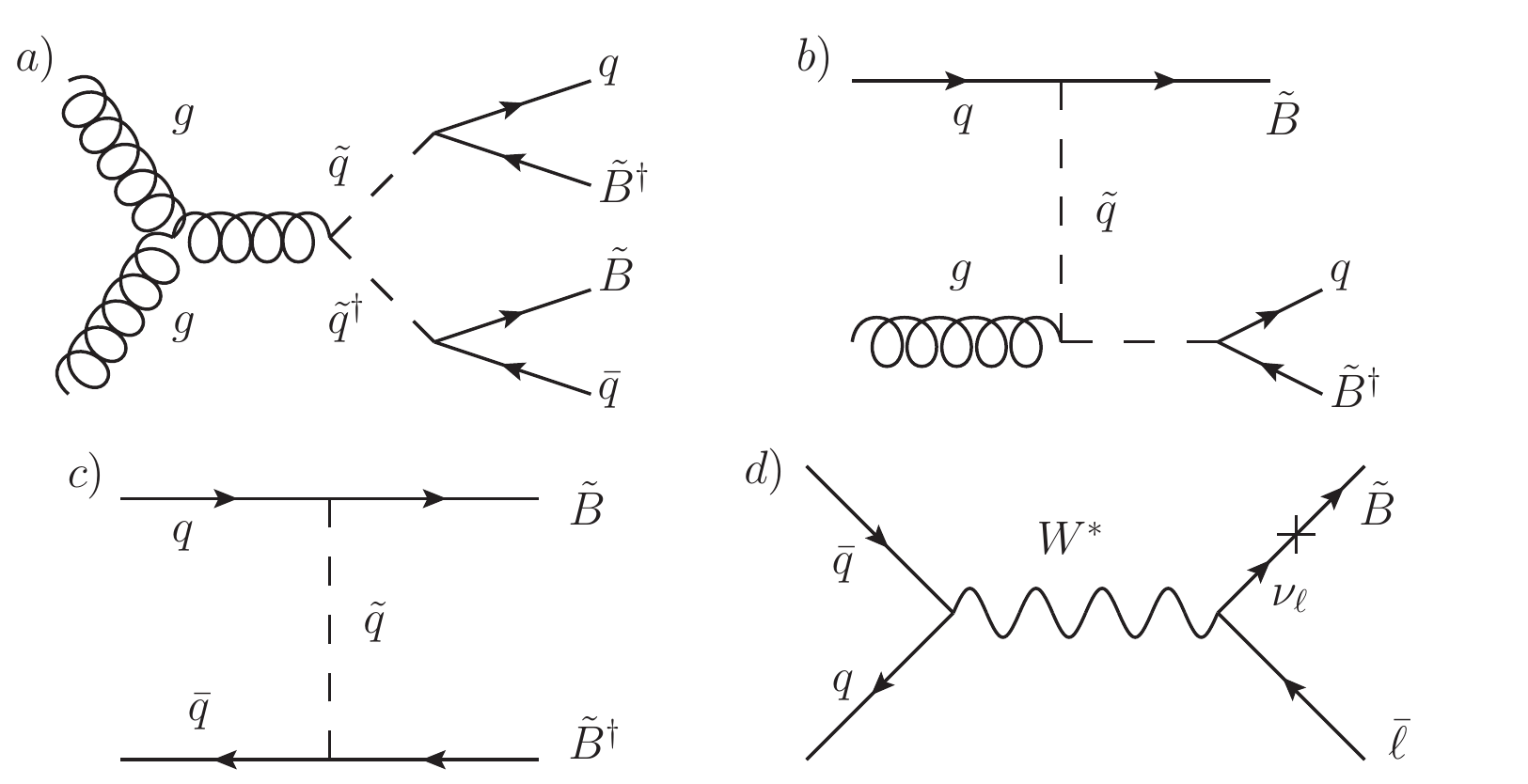}
\caption{Diagrams leading to $\bino$ production at the LHC. Leading contribution is through the squark-antisquark production and subsequent decays, shown in diagram (a).} \label{fig:binoprod}
\end{figure}

At hadron colliders, the \bivo can be produced in two main ways. The first mechanism is through off-shell electroweak bosons ($W$ and $Z$), just like any other heavy RH neutrino (see diagram (d) in Fig.~\ref{fig:binoprod}). However, this mode is strongly suppressed with the square of the neutrino-\bivo mixing angle, $\theta^2\sim (Yv/m_D)^2 \sim (v/\Lambda_M )^2 \sim \mathcal{O}(10^{-5})$. Thus, the cross section for this production mechanism would be too small to give an observable number of events at the LHC.

A much larger \bivo\!-production cross section is obtained via its interactions with colored particles, diagrams (a) - (c) in Fig.~\ref{fig:binoprod}. For instance, for a minimal supersymmetric SM scenario with degenerate squark masses $m_{\squark} \sim 1.5~\textrm{TeV}$, where stops and gluinos are decoupled, the squark pair production cross section (diagram (a) in Fig.~\ref{fig:binoprod}) at 13~TeV LHC is 2-3 fb~\cite{twiki}. Assuming a $100\%$ branching ratio for $\tilde{q}\to q\tilde{B}$, this would be the leading \bivo production mechanism at the LHC. Subleading contributions come from $g q \rightarrow \tilde{q} \tilde{B}$  and $ q \bar q \rightarrow \tilde{B} \tilde{B}$ (diagrams (b)-(c) in Fig.~\ref{fig:binoprod}), see \textit{e.g.}, Ref.~\cite{prospino} for cross section estimates. 

Once it is produced, the \bivo has four possible decay modes: \textbf{(i)} $\bino \rightarrow \tilde{G} \gamma$; \textbf{(ii)} $\bino \rightarrow W \ell $; \textbf{(iii)} $\bino \rightarrow Z \nu $; and \textbf{(iv)} $\bino \rightarrow h \nu$. The decay mode (i) is strongly suppressed with the Planck mass, $\Gamma (\bino \rightarrow \tilde{G} \gamma) \sim M_D^5/(M_{Pl}^2 m_{3/2}^2)\sim 10^{-8}~\textrm{eV}$. The rest of the decay modes are only suppressed with the neutrino-\bivo mixing and their branching ratios are approximately equal to $1/3$. This yields a total width  $\Gamma_{tot} \sim M_D Y^2 \sim M_D^3/\Lambda_M^2 \sim \mathcal{O} (500~\textrm{MeV})$, for $ M_D \sim 1~\textrm{TeV}$ and $\Lambda_M \sim 50~\rm{TeV}$. Thus, an important feature in this model is that the \bivo will decay promptly after being produced at the LHC, unlike in many other supersymmetric models where it leads to missing energy signatures (or displaced vertices, depending on its lifetime).

For the leading production mechanism via gluon fusion, assuming $\mathcal{B}(\squark \rightarrow \bino q) \sim 1 $, the final state will have two jets and two bi$\nu$os. Depending on whether the two bi$\nu$os decay via $W$, $Z$ or $h$, the signal at the detector will contain a certain combination of charged leptons, jets, and missing energy, \textit{e.g.},
\begin{equation}
 p p  \rightarrow 2j \; \; l^+ \; l^- \; 2J \, , 
\label{eq:LHC}
\end{equation}  
where $J$ stands for a wide jet produced in the decay of a boosted $W$. This signature is obtained when the two bi$\nu$os decay via a boosted $W$. There is no missing energy and all intermediate resonances ($\bino$, $W$, $\squark$) can be fully reconstructed. Thus, this would be the cleanest channel to search for the \bivo\!\!. The cross section at 13~TeV LHC can be obtained as 
\begin{eqnarray}
\sigma (g g \rightarrow \tilde{q} \tilde{q}) \times \mathcal{B}(\bino \rightarrow \ell W)^2 \mathcal{B}(W \rightarrow j j)^2 \simeq 0.16 \; \textrm{fb}\,, \nonumber 
\end{eqnarray}
where we have assumed that $\mathcal{B}(\squark \rightarrow \bino q) \sim 1 $, $\mathcal{B}(\bino \rightarrow W \ell) \sim 1/3 $, and we have used $ \sigma (g g \rightarrow \tilde{q} \tilde{q})\sim 3~\textrm{fb}$~\cite{twiki}. A characteristic feature of our model is that the branching ratios $\mathcal{B}(\bino \rightarrow \ell_i W)$ are fully determined by the flavor structure, which in turn is fixed by neutrino oscillation data. Thus, in case a positive signal is observed, further tests can be performed by comparing the signal rates for the processes shown in Eq.~\ref{eq:LHC} involving different charged leptons. 

Additional signatures involving more leptons and/or missing energy (for instance, when one or the two $W$ bosons decay leptonically) would be expected at a comparable rate. However, these would be more difficult to distinguish from the background as the intermediate resonances cannot be fully reconstructed. Furthermore, lepton number violating signatures at colliders are suppressed by the small \bivo Majorana mass and will not be observable.

Leptoquark searches may also apply in certain regions of the parameter space, for example, if the mass splitting between the $\tilde{q}$ and the $\bino$ is relatively small. ATLAS has the strongest constraints in this case: $\sigma(\mu\mu j j ) \lesssim 0.4$~fb for $m_{LQ} \sim 1$~TeV~\cite{Aad:2015caa}. In this model, an additional suppression with respect to the process in Eq.~\ref{eq:LHC} is obtained from requiring that both bi$\nu$os decay into muons, and therefore, $\sigma(pp \rightarrow 2j\; \mu \mu\; 2J) \sim (1/2)^2 \sigma(pp \rightarrow 2j \; \ell \ell \; 2J) \sim 0.04$~fb.  Thus, current limits are not strong enough, but future LHC data could further constrain this scenario.

\textit{Conclusions.} To summarize, we have argued that $U(1)_{R-L}$-symmetric supersymmetric models contain all the necessary ingredients to produce an ISS texture for the neutrino mass matrix. As the bino and the singlino have the appropriate quantum numbers, they can form a pseudo-Dirac pair (the \bivo\!\!) and play the role of right-handed neutrinos. No additional singlets are therefore needed. Furthermore, neutrino masses, as well as the bino and the singlino Majorana masses, are naturally suppressed since they explicitly violate $U(1)_{R-L}$. 

This model predicts the lightest neutrino to be massless. In order to explain the neutrino mass structure and respect the constraints from charged lepton flavor violating observables, the supersymmetric messenger scale should be $\Lambda_M\gtrsim 50~\rm{TeV}$, and the gravitino mass $m_{3/2}\sim \mathcal{O}({\rm keV})$. The collider phenomenology of the model has also been outlined.

\textit{Acknowledgments:} We thank Paddy Fox, Roni Harnik, Kiel Howe and Jacobo Lopez-Pavon for valuable conversations. SI thanks Ann Nelson for comments. Fermilab is operated by Fermi Research Alliance, LLC under Contract No. DE-AC02-07CH11359 with the United States Department of Energy. The work of SI was partly performed at the Aspen Center for Physics, which is supported by National Science Foundation grant PHY-1066293. This project has received funding from the European Union’s Horizon 2020 research and innovation programme under the Marie Sklodowska-Curie grant agreements No. 690575 and 674896.

\bibliography{ref}

\end{document}